\documentclass[12pt,a4paper]{article}
\usepackage{graphicx}
\usepackage{amsmath,amsthm,amsfonts,amssymb,mathrsfs}
\usepackage[latin1]{inputenc}
\newcommand{\myvec}[1]{{\boldsymbol#1}}
\theoremstyle{definition}
\newtheorem{rem}{Remark}

\begin{document}

\title{The uniqueness of inverse scattering problems, reciprocity
  principles, and nonradiating sources related to low-signature
  structures}

\author{
  Johan Helsing\thanks{Centre for Mathematical Sciences, Lund
    University, Box 118, 221 00 Lund, Sweden (johan.helsing@math.lth.se).}~~and
  Anders Karlsson\thanks{Electrical and Information Technology, Lund
    University, Box 118, 221 00 Lund, Sweden 
    (anders.karlsson@eit.lth.se).}}

\date{}
\maketitle

\begin{abstract}
  This paper is about perfectly electrically conducting structures
  designed to produce negligible scattered power when exposed to a
  time-harmonic plane electromagnetic wave. The structures feature
  cavities capable of concealing objects. Theoretical investigations
  of the properties of the structures combined with accurate numerical
  computations lead to three key findings: the first concerns the
  uniqueness of the solution to an inverse scattering problem, the
  second establishes a reciprocity relation for the far-field
  scattering amplitude, and the third reveals the existence of
  non-radiating sources that generate substantial electromagnetic
  fields near the source region. The results have applications in
  low-observable technology.
\end{abstract}

\section{Introduction}

A general physical object can be considered invisible to a given
incident electromagnetic wave if it is undetectable by a sensor from
any direction or distance. Much of the recent research on invisibility
has been focused on cloaking, as described in papers such
as~\cite{Fleury15, Vasquez09}. Cloaking involves covering an object
with a cloak that guides incident waves around the object without
producing a scattered wave. The search for suitable materials for
cloaking has led to extensive research in the area of
metamaterials~\cite{Fan21}. In microwave applications, such materials
exhibit properties not found in nature. Despite progress, significant
breakthroughs are still needed before cloaking can become a practical
technology.

A technology related to cloaking is stealth, also known as
low-observable technology. Its main goal is to minimize the
detectability of military structures such as aircraft, ships, and
submarines across various detection methods, including radar, infrared
sensors, and sonar. In radar applications, stealth techniques aim to
reduce the backscattering cross section, also known as the radar cross
section (RCS) \cite[Sec.~3.1]{Ahmad19}, of a structure, thereby making
it less visible to monostatic radar systems. This is achieved through
clever designs and the use of radar-absorbing materials
\cite{Ahmad19}. A structure with a small RCS is often described as
having a low radar signature. The term {\it low-signature structure},
as used in this paper, refers to a structure that produces minimal
scattered power in response to an incident wave. This indicates that
its total scattering cross section, rather than just the RCS, is
small.

This paper presents low-signature perfectly electrically conducting
(PEC) structures. As in many stealth applications the structures can
be fabricated from ordinary metals. They also function as cloaking
devices, featuring one or several cavities that can hide objects.
However, the designs of our structures differ from those used in
stealth and cloaking technologies. The cavities are located between
two horizontal infinitely thin PEC walls. When such a PEC structure is
illuminated by a time-harmonic linearly polarized electromagnetic
plane wave, the horizontal PEC walls enable the wave to pass by the
cavity, essentially without producing a scattered wave. Disadvantages
with our structures are that they need to be long in one spatial
direction and that the incident wave must be a transverse magnetic
(TM) wave with its magnetic field parallel to the horizontal walls.
The advantages are that the invisibility is far better than what can
be achieved by electromagnetic cloaking and that not only very low RCS
is achieved but also very low total scattering cross section.

A low-signature PEC structure, similar to the ones described in this
paper, is presented in \cite{Tretyakov09}. Its main purpose is to
reduce the signature of a cylindrical object by guiding incident waves
around it. In \cite{Tretyakov09}, this low-signature effect is
achieved only when the incident wave is a transverse electric (TE)
wave, with the electric field aligned parallel to the cylinder.

The first part of this paper deals with the design of the PEC
structures that can hide objects and have a low signature to an
incident TM plane wave in a wide or a narrow frequency band. Three
fundamental and unexpected findings are then presented regarding
low-signature structures.

The first finding concerns the non-uniqueness of the inverse
scattering problem for determining the boundary of PEC structures
based on measured scattering data.

The second finding, derived from the Lorentz reciprocity theorem,
states the following: suppose a structure is undetectable from any
direction or distance when illuminated by a TM plane wave traveling,
for example, in the positive $x$-direction. Then, place a sensor
either in the positive or negative $x$-direction, at a large distance
from the structure, and illuminate the structure by an arbitrary TM
wave generated by sources located anywhere outside the structure. Then
the sensor cannot detect the structure.

The third finding highlights the existence of nonradiating sources
that produce a significant electromagnetic field in a vicinity of the
source region.

All three findings emanate from physical assumptions and are supported
by numerical computations of scattered fields and scattering cross
sections. The computations are made using an integral equation method
\cite{HelsJiang25}, which offers the numerical precision necessary to
validate the findings. The numerical computations are made on
broadband low-signature structures and also on the narrowband
low-signature structures that were originally introduced in
\cite{HelsJiangKarl24} and \cite{HelsJiangKarl25}.

The rest of the paper is organized as follows. Section
\ref{formulation} presents the partial differential equations and
boundary conditions for the TM direct scattering problem. Section
\ref{lowsign} introduces the low-signature PEC structures. The three
findings of low-signature structures are presented in Section
\ref{features}. Concluding remarks are made in Section
\ref{conclusion}. A few notes on the numerical scheme
\cite{HelsJiang25} are collected in Appendix \ref{app:A}.

\section{The direct scattering problem}
\label{formulation}

\begin{figure}[t]
\centering
\includegraphics[scale=0.8]{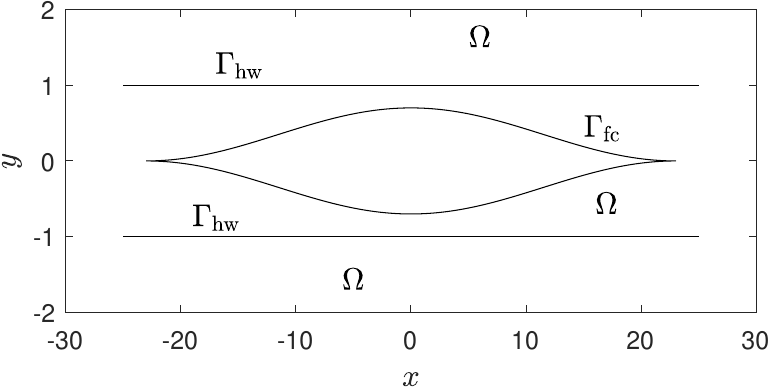}
\caption{\sf The geometric cross section of a broadband low-signature
  structure with a fusiform cavity with two cusps. The geometry is
  specified in \eqref{geombroad}. For the unit of length, see Remark
  \ref{rem:length}.}
\label{combgeom}
\end{figure}

\begin{figure}[h!]
\centering
  \includegraphics[scale=0.8]{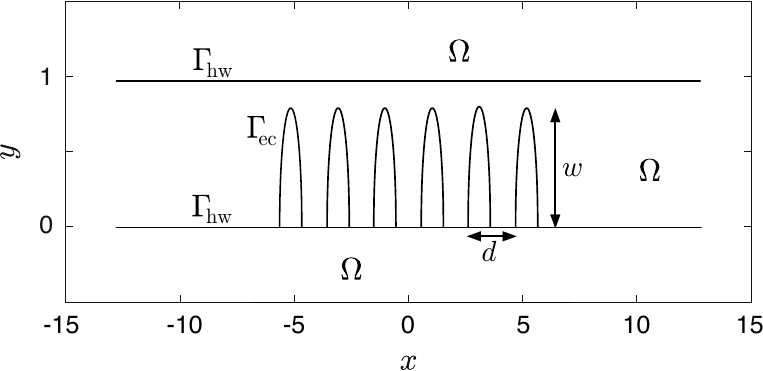}
  \caption{\sf The geometric cross section of a narrowband
    low-signature structure. The cavities are half ellipses with major
    axis $w$, minor axis $w/2$, and form a finite-periodic pattern
    with period $d$. The horizontal walls are specified in
    \eqref{geomnarrow}. For the unit of length, see Remark
    \ref{rem:length}.}
\label{geometrireciproc}
\end{figure}

The geometric cross sections of a broadband low-signature structure
and a narrowband low-signature structure are shown in
Figures~\ref{combgeom} and \ref{geometrireciproc}. Both structures are
translational invariant and have infinite extent along the
$z$-direction. Here $\Gamma_{\rm hw}$, $\Gamma_{\rm fc}$, and
$\Gamma_{\rm ec}$ denote the boundaries of horizontal walls, fusiform
cavities, and half-ellipse-shaped cavities, respectively. The
boundaries are all PEC. Throughout the paper, the notation $\Gamma$ is
used to represent the union $\Gamma_{\rm hw} \cup \Gamma_{\rm fc}$ and
the union $\Gamma_{\rm hw} \cup \Gamma_{\rm ec}$. The domain exterior
to $\Gamma$, denoted $\Omega$, contains air with relative permittivity
$\varepsilon_{\rm r}=1$. The domain exterior to the smallest rectangle
that circumscribes a given structure is denoted $\Omega_{\rm rec}$.
The horizontal walls are infinitely thin. The domain enclosed by
$\Gamma_{\rm fc}$ and the domains enclosed by $\Gamma_{\rm ec}$ are
cavities where objects can be hidden.

The structures are illuminated by a time-harmonic TM wave with angular
frequency $\omega$ and complex magnetic field
$\myvec H^{\rm in}(\myvec\rho)=H^{\rm in}(\myvec\rho)\hat{\myvec z}$,
where $\myvec\rho=x\hat{\myvec x}+y\hat{\myvec y}$, and
$\hat{\myvec x}$, $\hat{\myvec y}$, and $\hat{\myvec z}$ are the unit
vectors in a cartesian coordinate system. The transformation to the
time domain is
$\myvec H(\myvec\rho,t)=\Re\{\myvec H(\myvec\rho)e^{-{\rm i}\omega
  t}\}$. The total magnetic field
$\myvec H(\myvec\rho)=H(\myvec\rho)\hat{\myvec z}$ is the sum of the
incident and the scattered field
\begin{equation}
  H(\myvec\rho)=H^{\rm in}(\myvec\rho)+H^{\rm sc}(\myvec\rho),\quad
                \myvec\rho\in \Omega.
\label{decomp}
\end{equation}

The direct scattering problem is the exterior Neumann problem for
$H^{\rm sc}(\myvec\rho)$
\begin{align}
&(\nabla^2+k^2)H^{\rm sc}(\myvec\rho)=0, \quad \myvec\rho\in \Omega,\label{PDE1}\\
&\nu\cdot \nabla H^{\rm sc}(\myvec\rho)=-\nu\cdot \nabla H^{\rm in}(\myvec\rho), \quad  \myvec\rho\in \Gamma,\label{PDE2}\\
& H^{\rm sc}(\myvec\rho)\to \dfrac{e^{{\rm i}k\rho}}{\sqrt{\rho}}(F^{\rm sc}(\theta)+{\mathcal O}(\rho^{-1})),\quad \rho \to \infty.\label{PDE3}
\end{align}
Here, $k=\omega/c_0$ is the wavenumber, where $c_0$ is the speed of
light in vacuum, $\nu$ is the normal unit vector to $\Gamma$,
$\rho=\vert\myvec\rho\vert$, and $F^{\rm sc}(\theta)$ is the far-field
amplitude of the scattered field. The angle $\theta$ is the azimuthal
angle, measured from the $x$-axis.

The time average of the radiated power per unit length from a
structure is
\begin{equation}
P=\eta_0\int\limits_{-\pi}^\pi \vert F^{\rm sc}(\theta)\vert^2\,{\rm d}\theta,
\label{power}
\end{equation}
where $\eta_0=\sqrt{\mu_0/\varepsilon_0}$ is the wave impedance. When
the incident field is a plane wave, the (total) scattering cross
section is defined as
\begin{equation}
  \Sigma =\dfrac{\text{time average of scattered  power per unit length}}{\text{time average of incident power density}}.
\label{sigma}
\end{equation}
An incident TM plane wave has magnetic and electric fields
\begin{equation}\label{incidentObl}\begin{split}
&\myvec H^{\rm in}(\myvec\rho)=e^{{\rm i}\myvec k\cdot \myvec\rho}\hat{\myvec z},\\
&\myvec E^{\rm in}(\myvec\rho)=\eta_0e^{{\rm i}\myvec k\cdot \myvec\rho}\hat{\myvec z}\times \hat{\myvec k},
\end{split}
\end{equation}
where $\myvec k=k\hat{\myvec k}$ is the wave vector. The angle of
incidence is denoted $\alpha$ so that
\begin{equation}\label{alpha}
\hat{\myvec k}=(\cos\alpha,\sin\alpha,0).
\end{equation}
With incident field \eqref{incidentObl} then
\begin{equation}
\Sigma =\eta_0^{-1}P.
\end{equation}

\begin{rem}
  The unit of length is arbitrary and is omitted in the paper. It is
  understood that $\myvec \rho$, $k^{-1}$, and $\Sigma$ are expressed
  in the same unit of length and so are $x$ and $y$ in
  Figures~\ref{combgeom} and \ref{geometrireciproc}.
\label{rem:length}
\end{rem}

\section{Low-signature structures}
\label{lowsign}

With $\alpha=0$ in \eqref{alpha} then $\hat{\myvec k}=\hat{\myvec x}$
and \eqref{incidentObl} becomes
\begin{equation}\label{incidentperp}\begin{split}
&\myvec H^{\rm in}(\myvec\rho)=e^{{\rm i}kx}\hat{\myvec z},\\
&\myvec E^{\rm in}(\myvec\rho)=\eta_0e^{{\rm i}kx}\hat{\myvec y}.
\end{split}
\end{equation}
The structures in Figures~\ref{combgeom} and \ref{geometrireciproc}
are considered to have a low signature to~\eqref{incidentperp} if
$\Sigma $ is much less than the corresponding $\Sigma $ for the
structures without the horizontal walls.

A structure consisting only of the two PEC walls is called a {\it
  trivial invisible structure} since the incident wave
\eqref{incidentperp} then satisfies
$\nu\cdot \nabla H^{\rm in}(\myvec\rho)=0$ for $\myvec\rho\in \Gamma$.
Thus, (\ref{PDE1},\ref{PDE2},\ref{PDE3}) has the trivial solution
$H^{\rm sc}(\myvec\rho)=0$ for $\myvec\rho\in \Omega$ and, by that,
$\Sigma =0$ and the structure is entirely invisible to
\eqref{incidentperp}.

\subsection{Broadband  structures}

\begin{figure}[t]
\centering
\includegraphics[scale=.6]{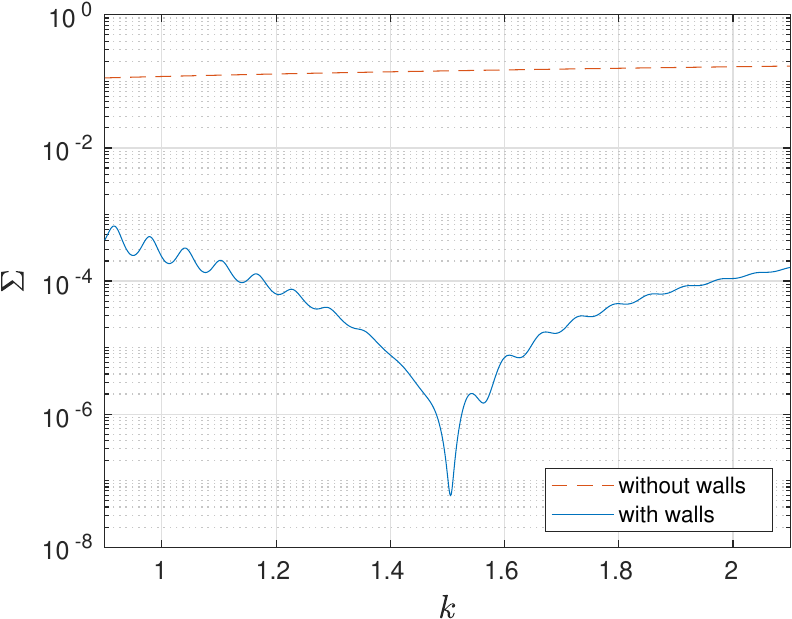}
\caption{\sf $\Sigma $ as a function of $k$ for the fusiform cavity in
  Figure \ref{combgeom}. Dashed red line: without horizontal
  walls. Solid blue line: with horizontal walls. The minimum is at
  $k=1.505395$.}
\label{sigmaperp}
\end{figure}

In the geometric cross section $z=0$, the horizontal walls and the
fusiform cavity of the example structure in Figure~\ref{combgeom} have
the parameterization $\myvec\rho(s) = (x(s), y(s),0)$, where
\begin{equation}\label{geombroad}\begin{split}
&\myvec\rho(s)=(25 s,\pm 1,0),\quad s\in [-1,1],\\
&\myvec\rho(s)=(23s,\pm0.7\cos^2(\pi s/2),0),\quad s\in [-1,1].
\end{split}
\end{equation}
The low signature resulting from \eqref{incidentperp} is illustrated
by the numerical computations in Figure~\ref{sigmaperp}. Across a
broad frequency range, the fusiform cavity positioned between
horizontal walls demonstrates a significantly smaller $\Sigma$
compared to the same cavity without the horizontal walls. This is
quite remarkable and the reason for this low signature is as follows:

The horizontal walls function as a planar waveguide that supports the
propagation of transverse electromagnetic (TEM) waves. When the
fusiform cavity is situated between these walls the waveguide
effectively splits into two separate guides -- one above and one below
the cavity. The fusiform shape allows the TEM wave to pass above and
below the cavity with minimal reflection. After passing the cavity,
the two waves recombine to form a TEM wave with nearly the same
amplitude and phase as the incident wave.

\begin{figure}[h!]
\centering
\includegraphics[scale=.51]{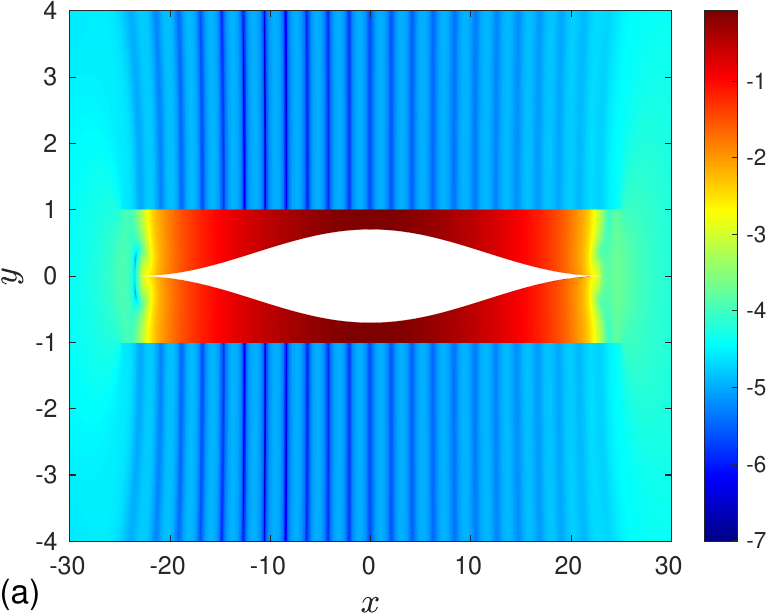}
\includegraphics[scale=.51]{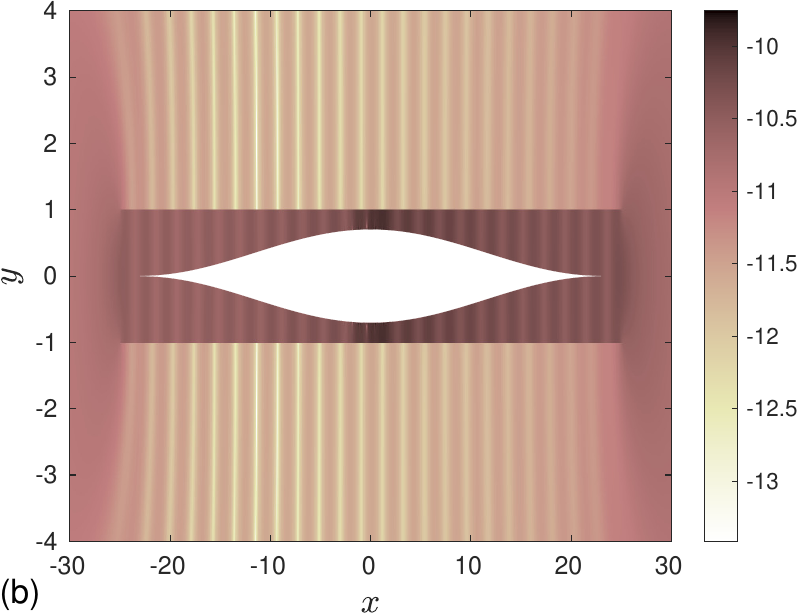}
\caption{\sf The structure in Figure~\ref{combgeom} illuminated by
  \eqref{incidentperp} at $k=1.505395$; (a) $\log_{10}$ of
  $\vert H^{\rm sc}(\myvec\rho)\vert$; (b) $\log_{10}$ of estimated
  absolute pointwise error in $H^{\rm sc}(\myvec\rho)$.}
\label{fusiformfields}
\end{figure}

The minimum of $\Sigma$ of the blue graph in Figure~\ref{sigmaperp} is
at $k=1.505395$. A small $\Sigma$ does not necessarily imply that
$H^{\rm sc}(\myvec\rho)$ remains small in the near field. However, the
field image of $\log_{10}$ of $\vert H^{\rm sc}(\myvec\rho)\vert$ in
Figure~\ref{fusiformfields}, along with $\log_{10}$ of the estimated
absolute pointwise error, clearly indicate that the scattered magnetic
near field is negligible within $\Omega_{\rm rec}$. The $x$- and
$y$-axes are scaled differently to provide a good field resolution
between the walls.

\begin{figure}[h!]
\centering
\includegraphics[scale=.6]{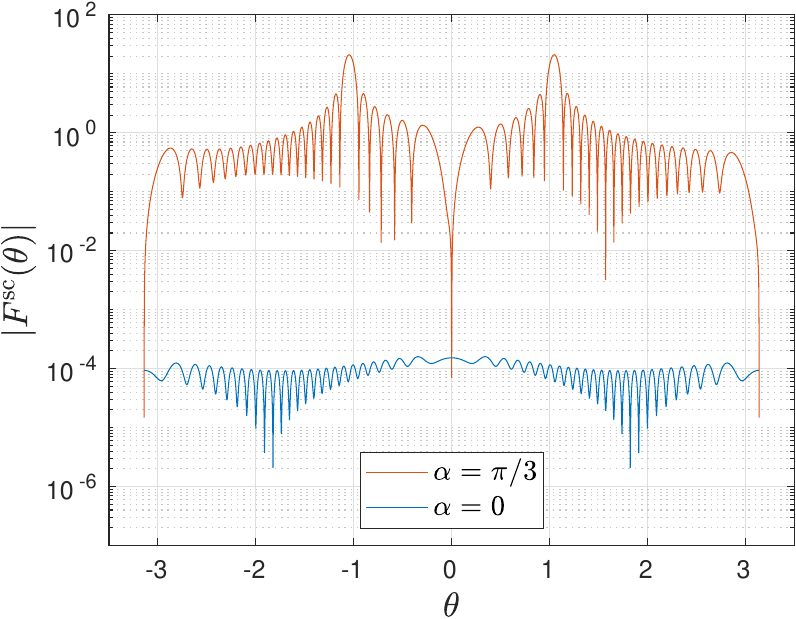}
\caption{\sf The structure in Figure~\ref{combgeom} illuminated by
  \eqref{incidentObl} at $k=1.505395$. Red line: $\log_{10}$ of
  $\vert F^{\rm sc}(\theta)\vert$ at oblique incidence with
  $\alpha=\pi/3$; Blue line: $\log_{10}$ of
  $\vert F^{\rm sc}(\theta)\vert$ at perpendicular incidence
  ($\alpha=0$); }
\label{combined4}
\end{figure}

A final numerical example demonstrating the broadband low signature is
presented in Figure~\ref{combined4}. It depicts the magnitude of
$\vert F^{\rm sc}(\theta)\vert $ for $-\pi < \theta < \pi$, with the
angles of incidence $\alpha=0$ and $\alpha=\pi/3$ at $k=1.505395$. One
may notice that $\vert F^{\rm sc}(0)\vert$, and
$\vert F^{\rm sc}(\pm \pi)\vert$ are very small when
$\alpha=\pi/3$. This phenomenon is due to reciprocity and is explained
in Section~\ref{reciprocity} below.

\subsection{Narrowband structures}

The example narrowband low-signature structure in
Figure~\ref{geometrireciproc} has a finite-periodic set of six PEC
half-ellipse-shaped cavities between the horizontal walls. The shape
and number of cavities can vary, but they must be identical and
equally spaced to achieve low signature. The low-signature property of
such narrowband structures can be explained by waveguide theory
\cite[Sec.~IIB]{HelsJiangKarl24}.

\begin{figure}[t]
\includegraphics[scale=.526]{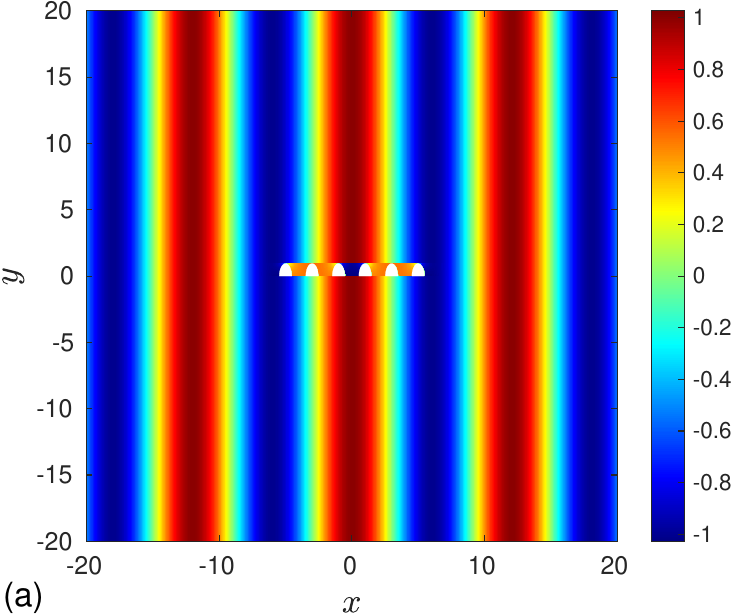}
\includegraphics[scale=.526]{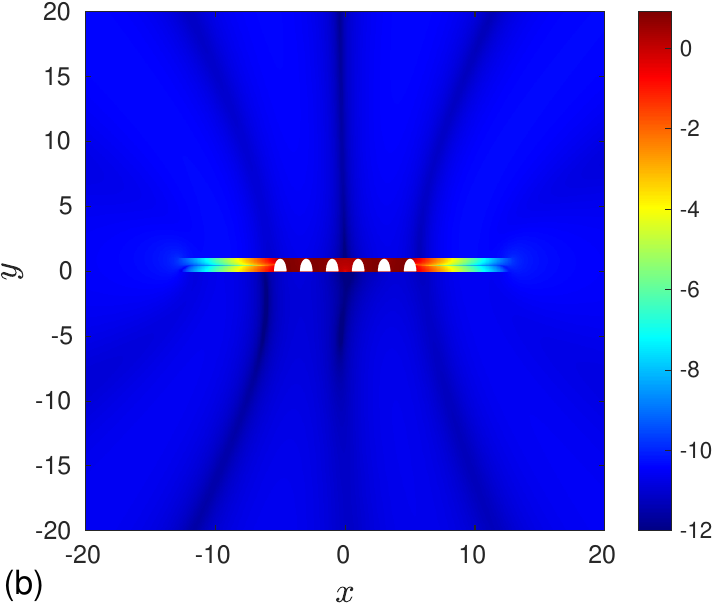}
\includegraphics[scale=.526]{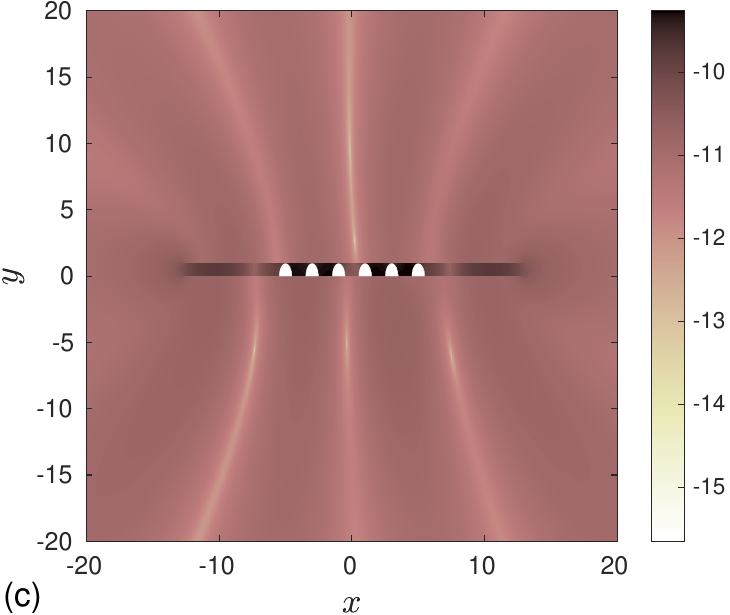}\hspace*{3mm}
\includegraphics[scale=.5]{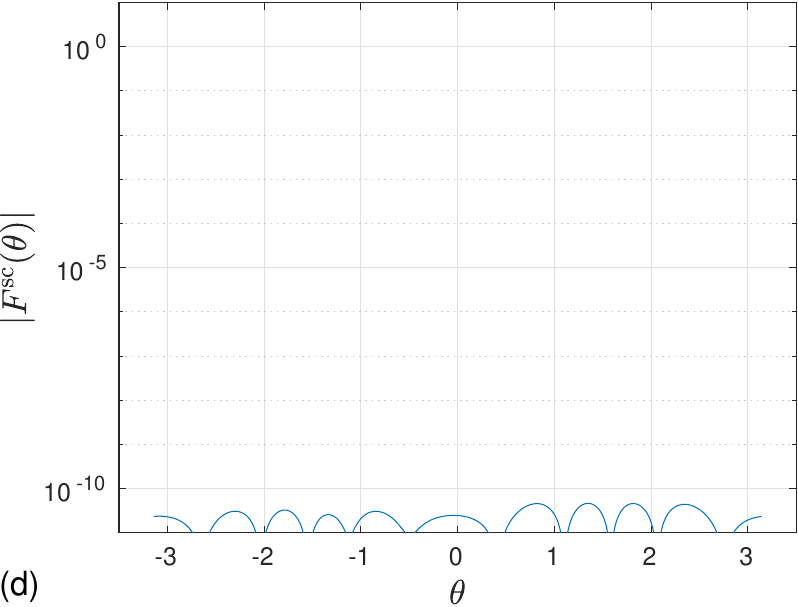}
\caption{\sf The plane wave \eqref{incidentperp} incident on the
  structure in Figure~\ref{geometrireciproc} with $k=0.523371641652$,
  and with $d=2$ and $w=0.939225885952$ in \eqref{geomnarrow}. (a)
  real part of $H(\myvec\rho)$; (b) $\log_{10}$ of
  $\vert H^{\rm sc}(\myvec\rho)\vert$; (c) $\log_{10}$ of estimated
  absolute pointwise error in $\vert H^{\rm sc}(\myvec\rho)\vert$; (d)
  $\vert F^{\rm sc}(\theta)\vert$.}
\label{plan0}
\end{figure}

The horizontal walls in Figure~\ref{geometrireciproc} are given by
\begin{equation}\label{geomnarrow}
\myvec\rho(s)=((7.3+2.5d+w/2)s,0.5\pm0.5,0),\quad s\in [-1,1].
\end{equation}
The half ellipses have period $d=2$, major axes $w$, and minor axes
$w/2$, with $w=0.939225885952$. Figure~\ref{plan0}(d) shows that
$F^{\rm sc}(\theta)$ is negligible for all $\theta$ when
$k=0.523371641653$ in \eqref{incidentperp}. Consequently,
by~\eqref{sigma} and~\eqref{power}, the structure has an almost zero
signature to \eqref{incidentperp} at this wavenumber. Furthermore,
Figure~\ref{plan0} shows that the scattered magnetic near field is
almost zero within $\Omega_{\rm rec}$. It should be noted, though,
that the low signature is maintained only within a very narrow
frequency band. Narrowband structures of the type shown in
Figure~\ref{geometrireciproc} are mainly effective in applications
where the wave frequency is precisely known.

\section{ Features of low-signature structures}
\label{features}

Three somewhat surprising features of low-signature structures are now
presented.

\subsection{The inverse scattering problem}
\label{inversescattering} 

Inverse scattering has been a highly active area of research for well
over 50 years \cite{ColtonKress18}. Inverse electromagnetic scattering
focuses on finding the properties of an object using data collected
from direct scattering experiments. The inverse scattering problem
addressed here is to determine the shape of the boundary of a
two-dimensional (2D) PEC structure from measured values of the
scattered electric or magnetic field due to an incident TM or TE wave.
If the scattered magnetic or electric fields are measured with finite
sensitivity in $\Omega_{\rm rec}$, then the following holds:

\emph{The inverse scattering 2D problem of determining the shape of a
  PEC structure is not guaranteed to have a unique solution if the
  incident waves are restricted to be 2D single-frequency TM or TE
  waves generated by sources in $\Omega_{\rm rec}$.}

\begin{figure}[t]
\includegraphics[scale=.526]{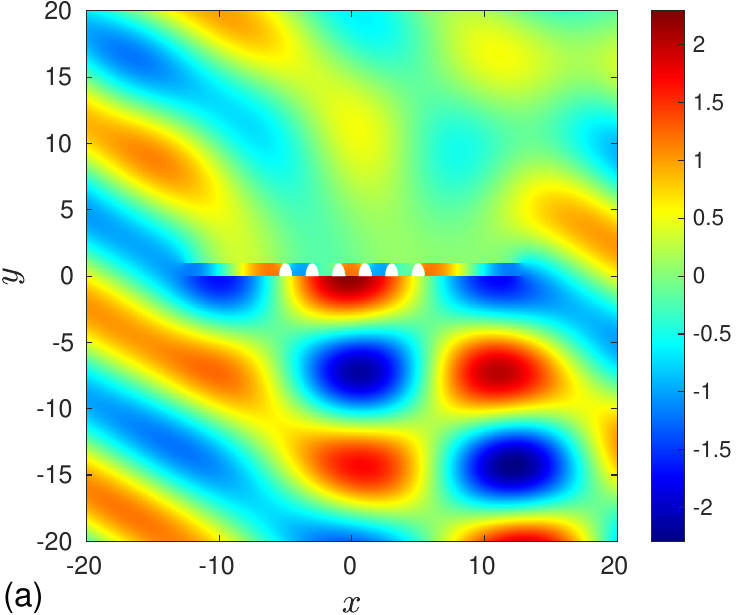}
\includegraphics[scale=.526]{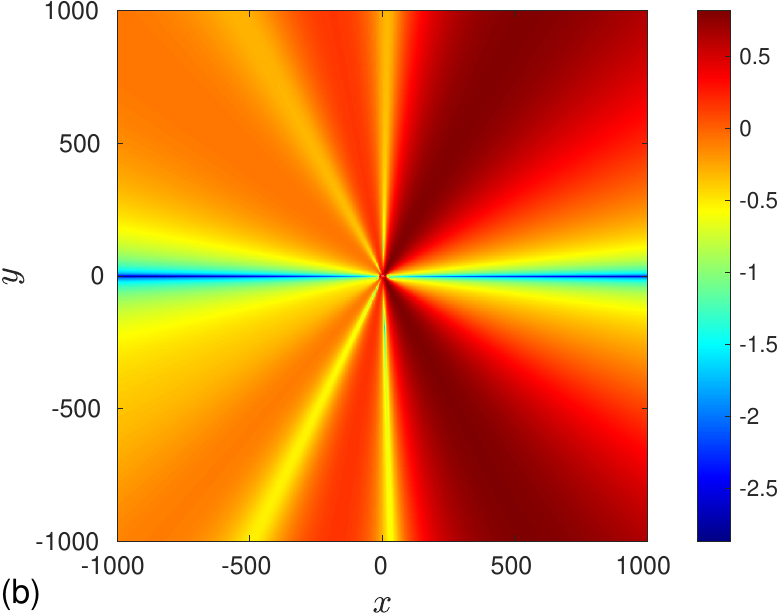}
\includegraphics[scale=.51]{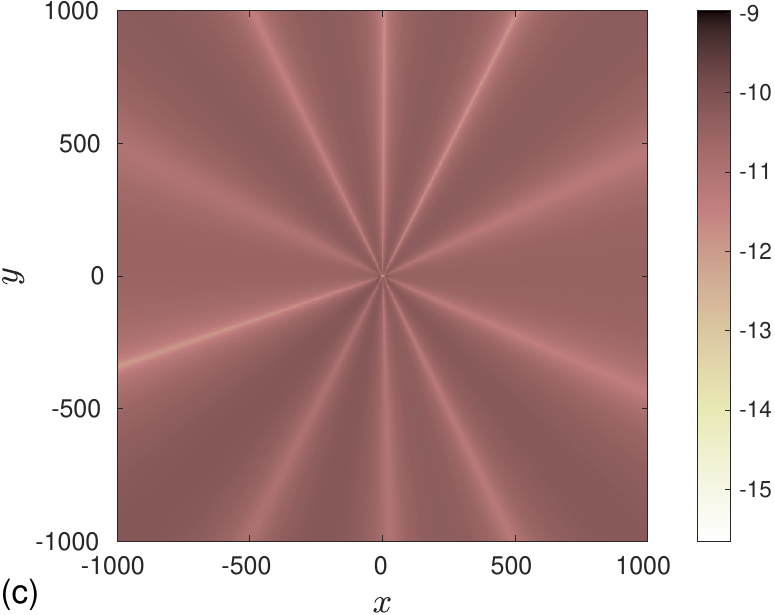}\hspace*{3mm}
\includegraphics[scale=.5]{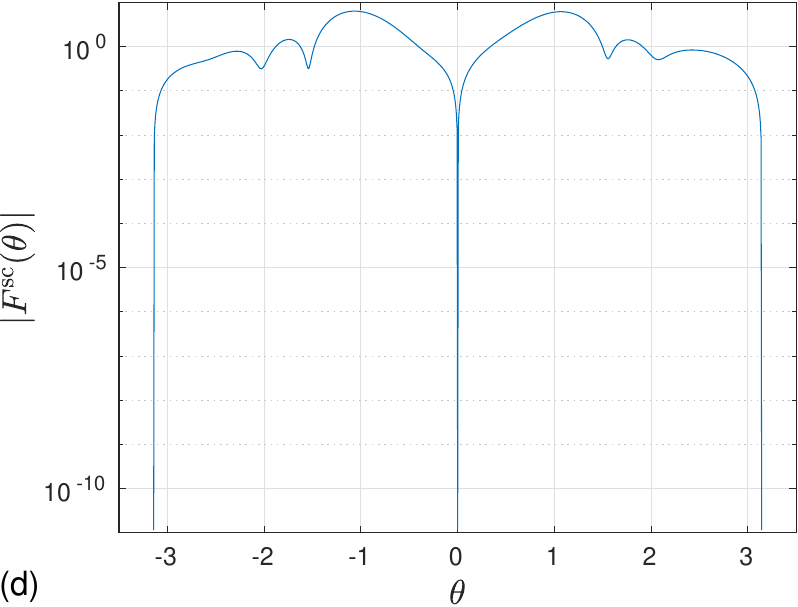}
\caption{\sf The plane wave \eqref{incidentObl} with $\alpha=\pi/3$
  incident on the structure in Figure~\ref{geometrireciproc}: (a) real
  part of $H(\myvec\rho)$. (b) $\log_{10}$ of
  $\vert H^{\rm sc}(\myvec\rho)\vert\sqrt{\rho}$; (c) $\log_{10}$ of
  estimated absolute pointwise error in
  $\vert H^{\rm sc}(\myvec\rho)\vert\sqrt{\rho}$; (d)
  $\vert F^{\rm sc}(\theta)\vert$.}
\label{plan60}
\end{figure}

\begin{figure}[t]
\includegraphics[scale=.526]{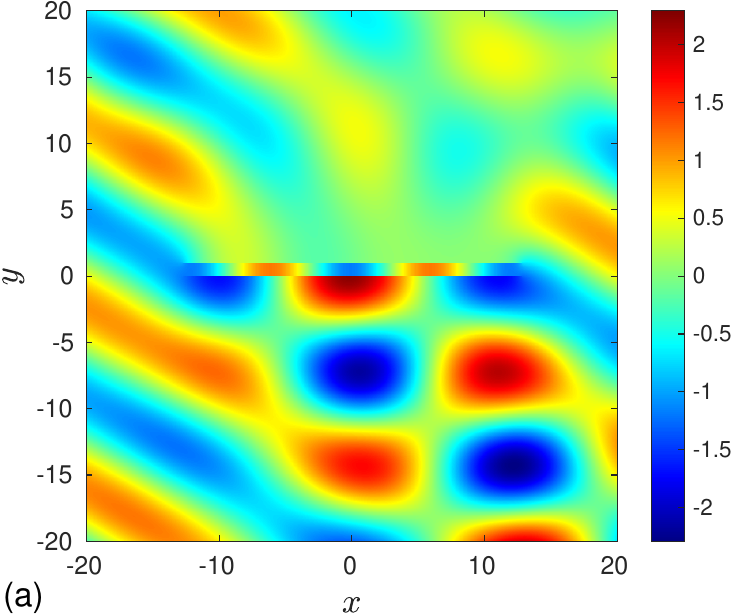}
\includegraphics[scale=.526]{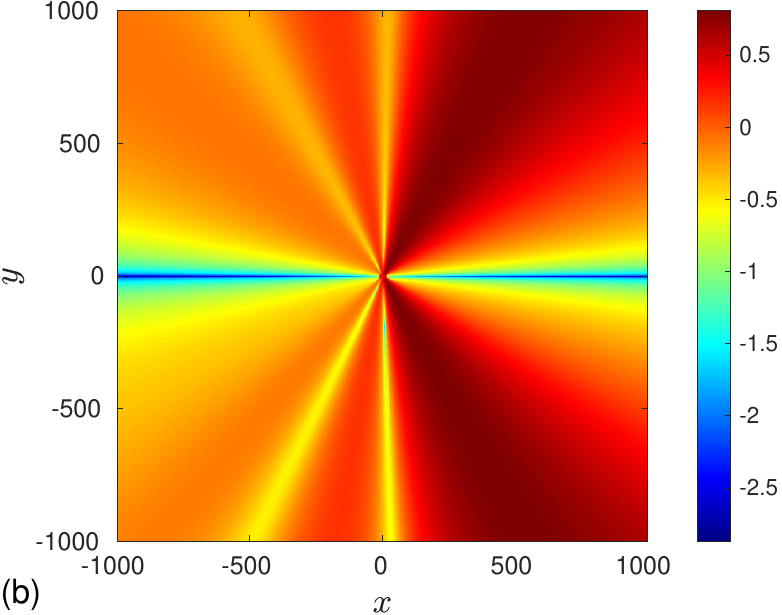}
\includegraphics[scale=.51]{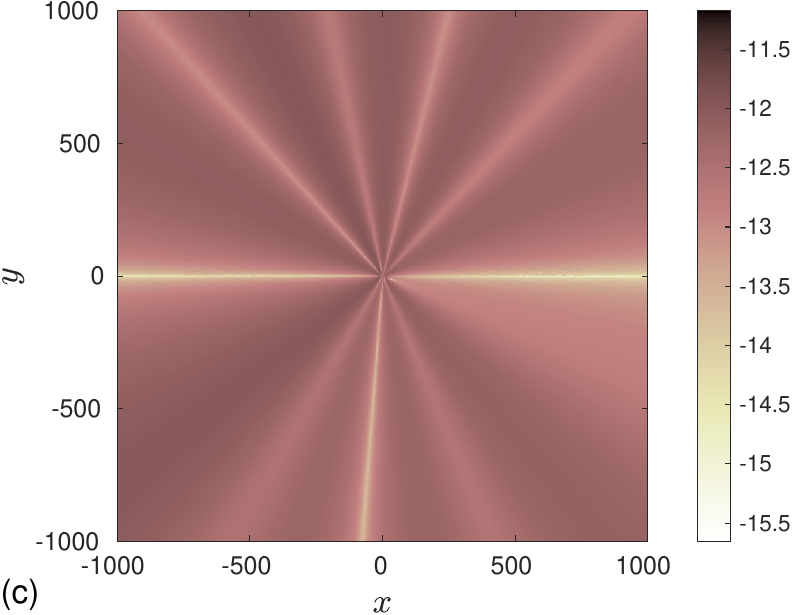}\hspace*{3mm}
\includegraphics[scale=.5]{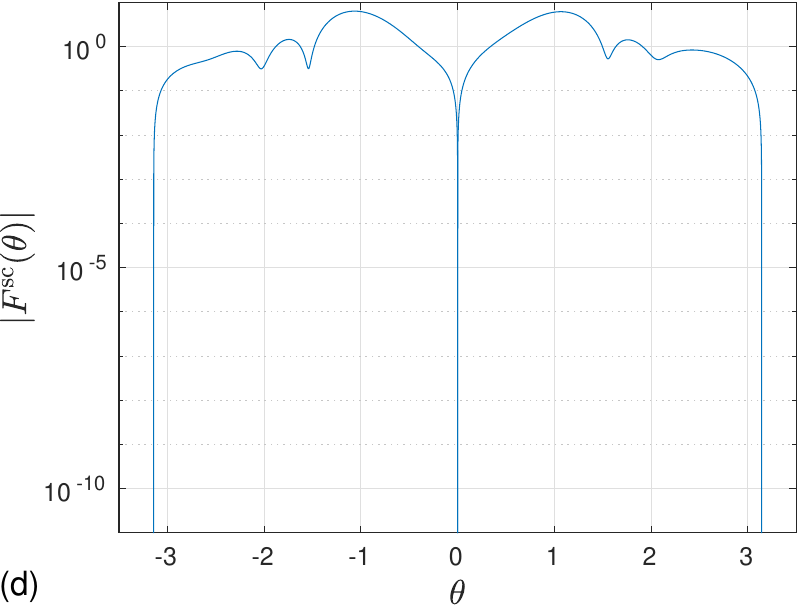}
\caption{\sf The plane wave \eqref{incidentObl} with $\alpha=\pi/3$
  incident on the structure in Figure~\ref{geometrireciproc} without
  the half-ellipse-shaped cavities: (a) real part of $H(\myvec\rho)$.
  (b) $\log_{10}$ of $\vert H^{\rm sc}(\myvec\rho)\vert$; (c)
  $\log_{10}$ of estimated absolute pointwise error in
  $\vert H^{\rm sc}(\myvec\rho)\vert$; (d)
  $\vert F^{\rm sc}(\theta)\vert$.}
\label{nobarr60}
\end{figure}

\begin{figure}[t]
\centering
\includegraphics[scale=.5]{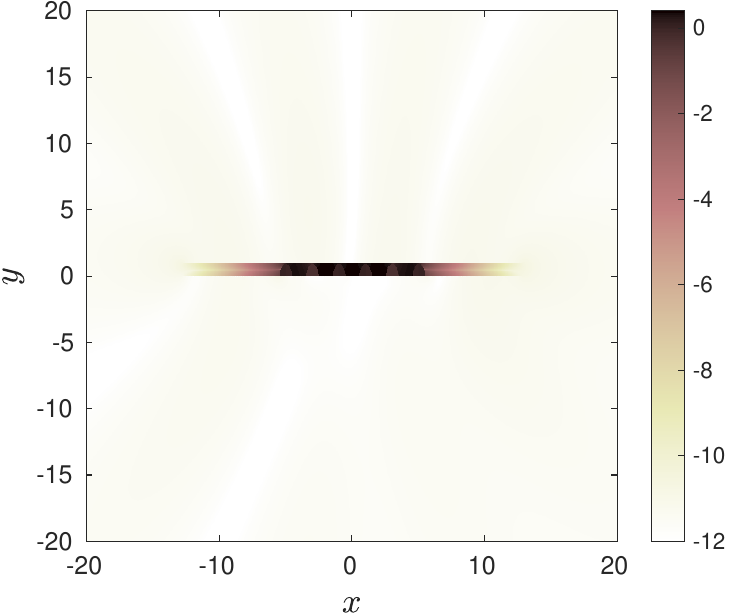}
\caption{\sf $\log_{10}$ of
  $\vert H_{\rm A}(\myvec\rho)-H_{\rm B}(\myvec\rho)\vert$. The real
  parts of $H_{\rm A}(\myvec\rho)$ and $H_{\rm B}(\myvec\rho)$ are
  shown in Figures~\ref{plan60}(a) and \ref{nobarr60}(a),
  respectively.}
\label{outside}
\end{figure}

We now confirm this statement. Let structure~{\sf A} be the structure
in Figure~\ref{geometrireciproc}, with $d=2$ and $w=0.939225885952$ in
\eqref{geomnarrow}, and let structure~{\sf B} be the corresponding
trivial structure, that is the structure with only the horizontal
walls. Let $k=0.523371641653$ and the incident field be an arbitrary
2D wave generated by sources in $\Omega_{\rm rec}$. Then the two
structures have the same scattered electromagnetic fields in
$\Omega_{\rm rec}$. Waveguide theory provides an explanation:

First, let the incident wave be TM. For both structures, the incident
wave gives rise to two TEM waves between the horizontal walls, one at
the opening to the left traveling in the positive $x$-direction, and
one at the opening to the right, traveling in the negative
$x$-direction. The cavities in structure~{\sf A} are designed so that
the TEM waves are transmitted between the openings in the same way as
they are transmitted between the openings in structure~{\sf B}. Thus,
the contribution by the TEM waves to $H^{ \rm sc}(\myvec\rho)$ for
$\myvec\rho\in \Omega_{\rm rec}$ must be the same for the two
structures. The incident wave also excites the planar waveguide TM$_n$
modes with $n>0$ \cite[Eq.~(11)]{HelsJiangKarl24} at the openings, but
these modes attenuate fast and have a negligible amplitude at the
cavities.

Next, let the incident wave be TE. At the openings it can only excite
the planar waveguide TE$_{n}$ modes with $n>0$ and at
$k=0.523371641653$ they attenuate fast enough to have a negligible
amplitude at the cavities. Thus the scattered fields in
$\Omega_{\rm rec}$ are not affected by the cavities and by that the
scattered fields from structure~{\sf A} is the same as those from
structure~{\sf B}.

To further verify the statement we present two numerical examples, one
with an incident TM wave and one with an incident TE wave.

Let the incident wave be \eqref{incidentObl} with $k=0.523371641653$,
and $\alpha=\pi/3$ in \eqref{alpha}. The resulting field images of the
real part of $H(\myvec\rho)$ are shown in Figure~\ref{plan60}(a) for
structure~{\sf A} and in Figure~\ref{nobarr60}(a) for structure~{\sf
  B}. In $\Omega_{\rm rec}$ the field image of Figure~\ref{plan60}(a)
is indistinguishable from the image of Figure~\ref{nobarr60}(a). The
similarity is confirmed by Figure~\ref{outside} which shows
$\vert H_{\sf A}(\myvec\rho)-H_{\sf B}(\myvec\rho)\vert$ in the square
$-20<x,y<20$. In the part of the square belonging to
$\Omega_{\rm rec}$ the difference is less than $10^{-9}$.

\begin{figure}[t]
\centering
\includegraphics[scale=.5]{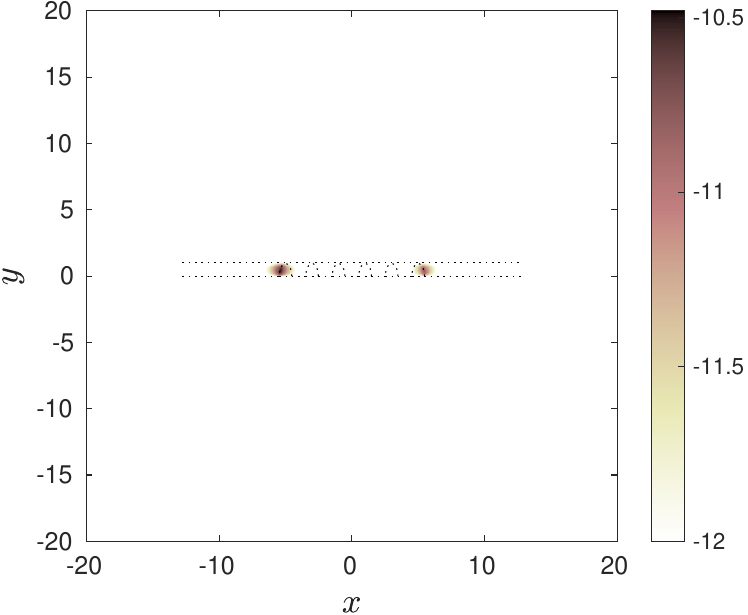}
\caption{\sf $\log_{10}$ of
  $\vert E_{\rm A}(\myvec\rho)-E_{\rm B}(\myvec\rho)\vert$ for the TE
  wave \eqref{incidentTE} with $\alpha=\pi/3$ and $k=0.523371641653$
  incident on the structure~{\sf A} and~{\sf B}, where~{\sf A} is the
  structure in Figure~\ref{geometrireciproc} with cavities and~{\sf B}
  is the same structure without cavities.}
\label{nobarr60D}
\end{figure}

Next, let the incident wave be the TE wave
\begin{equation}\label{incidentTE}\begin{split}
&\myvec H^{\rm in}(\myvec\rho)=e^{{\rm i}\myvec k\cdot \myvec\rho}\hat{\myvec k}\times \hat{\myvec z},\\
&\myvec E^{\rm in}(\myvec\rho)=\eta_0 e^{{\rm i}\myvec k\cdot \myvec\rho}\hat{\myvec z}.
\end{split}
\end{equation}
The structures~{\sf A} and~{\sf B}, the angle of incidence $\alpha$,
and the wavenumber $k$ remain unchanged from those used for the TM
wave above. The difference
$\vert E_{\sf A}(\myvec\rho)-E_{\sf B}(\myvec\rho)\vert$ is shown in
Figure~\ref{nobarr60D}. Note that TE waves are polarized such that
$\myvec E(\myvec\rho)=E(\myvec\rho)\hat{\myvec z}$, where
$E(\myvec\rho)$ satisfies the exterior Dirichlet problem. This problem
is given by (\ref{decomp},\ref{PDE1},\ref{PDE3}), with $H$ replaced by
$E$, and the boundary condition
$ E^{\rm sc}(\myvec\rho)=-E^{\rm in}(\myvec\rho)$ for
$\myvec\rho\in \Gamma$.

Figures~\ref{outside} and \ref{nobarr60D} show that
$\vert H_{\sf A}(\myvec\rho)- H_{\sf B}(\myvec\rho)\vert\approx 0$ and
$\vert E_{\sf A}(\myvec\rho)- E_{\sf B}(\myvec\rho)\vert\approx 0$ for
$\myvec\rho\in\Omega_{\rm rec}$. This means that the inverse problem
of reconstructing the boundary of a PEC structure from experimental
data, obtained with plane waves at a single wavenumber, is not
necessarily uniquely solvable. This conclusion applies not only to
incident plane waves, but to any 2D TM or TE incident wave with
sources located in $\Omega_{\rm rec}$.

\subsection{Reciprocity applied to invisible structures}
\label{reciprocity}
 
Consider a structure similar to the one in
Figure~\ref{geometrireciproc} that is invisible to
\eqref{incidentperp} at a wavenumber $k=k_{\rm inv}$. Using the
Lorentz reciprocity theorem, \cite[Sec.~1.3.5]{Kristensson16}, one can
prove that $F^{\rm sc}(0)\approx 0$ and $F^{\rm sc}(\pi)\approx 0$
when $k=k_{\rm inv}$ and the incident wave is an arbitrary TM wave
generated by a source in $\Omega_{\rm rec}$.

\begin{figure}[t]
\includegraphics[scale=.526]{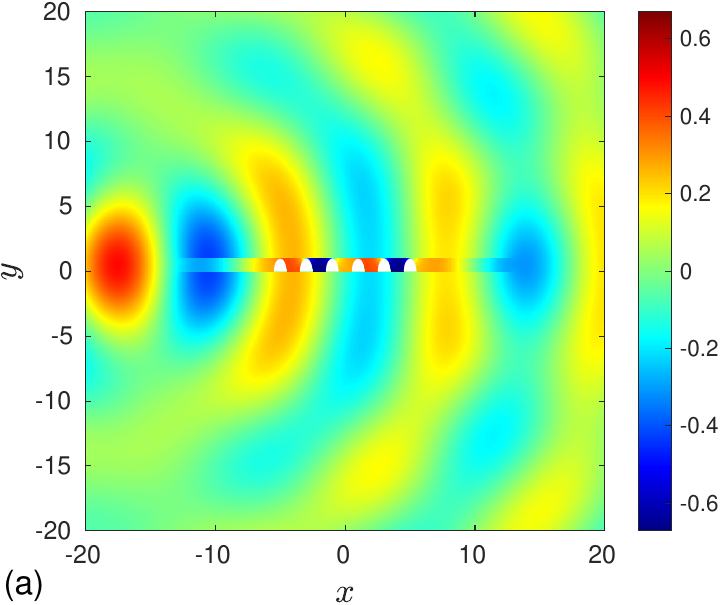}
\includegraphics[scale=.526]{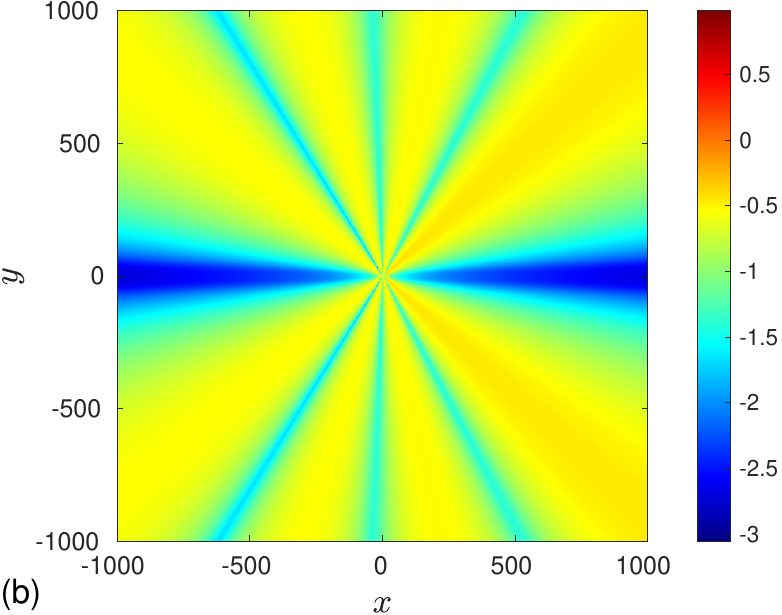}
\includegraphics[scale=.51]{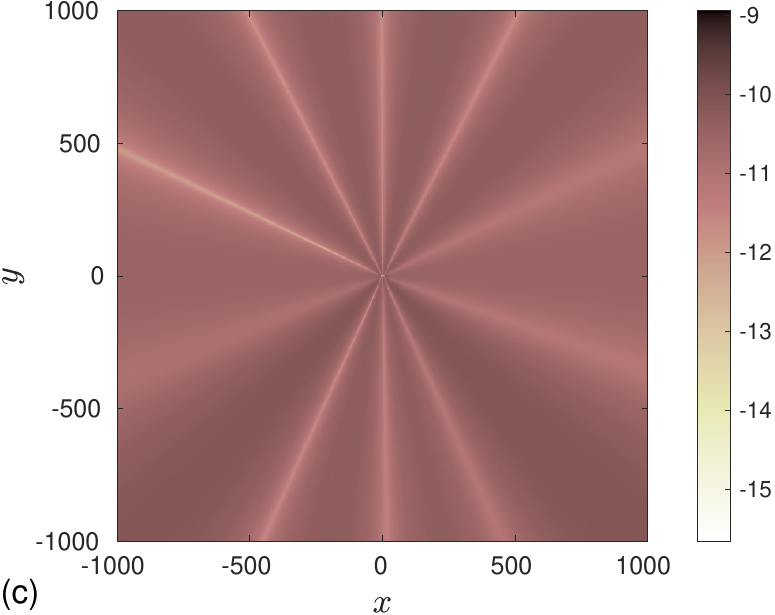}\hspace*{3mm}
\includegraphics[scale=.5]{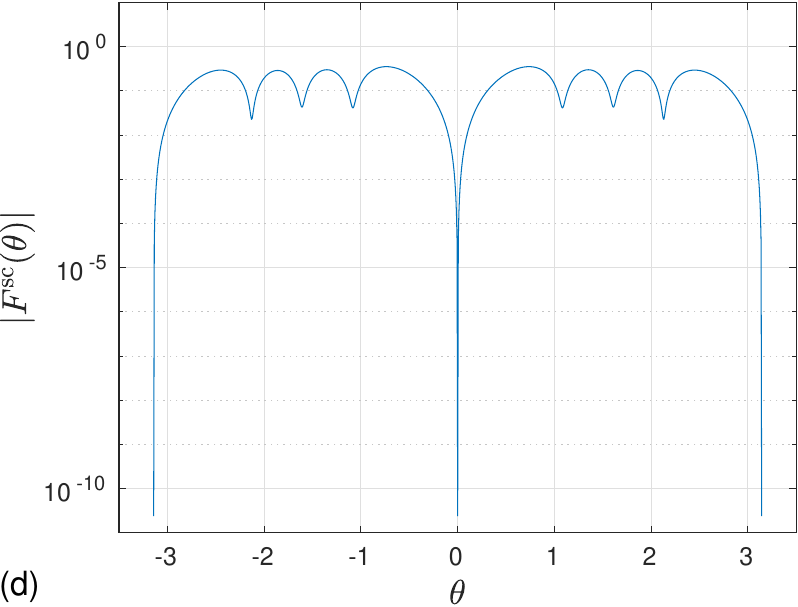}
\caption{\sf Field images for the structure in
  Figure~\ref{geometrireciproc} with incident field \eqref{dipol2},
  $\myvec\rho_0=(-14,0.5,0)$, and $k=0.523371641653$: (a) real part of
  $H(\myvec\rho)$; (b) $\log_{10}$ of
  $\vert H^{\rm sc}(\myvec\rho)\vert\sqrt{\rho}$; (c) $\log_{10}$ of
  estimated absolute pointwise error in
  $\vert H^{\rm sc}(\myvec\rho)\vert\sqrt{\rho}$; (d)
  $\vert F^{\rm sc}(\theta)\vert$.}
\label{dipol360}
\end{figure}

\subsubsection{Numerical examples for reciprocity}

Figures~\ref{plan60}(d) and \ref{nobarr60}(d) show that, in accordance
with reciprocity, $F^{\rm sc}(0)\approx 0$ and
$F^{\rm sc}(\pi)\approx 0$ for \eqref{incidentObl} with
$\alpha=\pi/3$. The reciprocity is also confirmed in
Figure~\ref{combined4} where $F^{\rm sc}(0)$ and $F^{\rm sc}(\pm \pi)$
are very small for the broadband structure when $\alpha=\pi/3$ and
$k=1.505395$.

As a final verification of reciprocity, we consider the incident wave
generated by a 2D electric dipole with its current directed along
$\hat{\myvec{y}}$. The incident magnetic field from this dipole is
\begin{equation}\label{dipol2}
\myvec H^{\rm in}(\myvec\rho)=-\sqrt{\dfrac{\pi k}{2}}\dfrac{(\myvec\rho-\myvec\rho_0)\times\hat{\myvec y}}{\vert \myvec\rho-\myvec\rho_0\vert}H_1^{(1)}(k\vert \myvec\rho-\myvec\rho_0\vert ).
\end{equation}
Here $\myvec{\rho}_0=x_0\hat{\myvec x}+y_0\hat{\myvec y}$ is the
position of the dipol and $H_1^{(1)}$ is the first order Hankel
function of the first kind. The incident field \eqref{dipol2} is
normalized such that the far-field amplitude, also called the
radiation pattern, of the dipole in free space is
\begin{equation}\label{dipolfarfield}
\vert F(\theta)\vert=\vert \sin\theta\vert.
\end{equation}

Let the structure be the same as in Figure~\ref{plan0}, and let
$k=0.523371641653$. With $\myvec\rho_0=(-14,0.5,0)$ the real part of
$H(\myvec\rho)$, $\log_{10}$ of $\vert H^{\rm sc}(\myvec\rho)\vert$,
$\log_{10}$ of estimated absolute pointwise error in
$\vert H^{\rm sc}(\myvec\rho)\vert$, and
$\vert F^{\rm sc}(\theta)\vert$ are as in Figure~\ref{dipol360}. It is
clear from Figure~\ref{dipol360}(d) that $F^{\rm sc}(0)\approx 0$ and
$F^{\rm sc}(\pi)\approx 0$. The singularity of the incident field at
$\myvec \rho=\myvec \rho_0$ is not visible in Figure~\ref{dipol360}(a)
because the real part of
$H_1^{(1)}(k\vert \myvec \rho-\myvec \rho_0\vert )$ is finite
everywhere.

\subsection{Nonradiating sources}

In the example shown in Figure \ref{plan0}, the sources for
$H^{\rm sc}(\myvec\rho)$ are the surface currents on
$\Gamma$. Figure~\ref{plan0}(d) shows that the radiated power
\eqref{power} from the narrowband structure is negligible at
$k=0.523371641652$ since $F^{\rm sc}(\theta)\approx 0$ for all
$\theta$. According to the definitions of nonradiating sources,
\cite[p.~275]{Gbur03}, \cite[p.~1]{LiWang24}, \cite[p.~2]{KimWolf86},
the surface currents on $\Gamma$ thus form a nonradiating source. It
is a well-established result, \cite[Thm.~3.2]{Friedlander73}, that the
field radiated from a nonradiating 3D source is zero everywhere
outside a sphere circumscribing the source. The theorem also applies
in two dimensions if the sphere is replaced by a circle. This is in
accordance with Figure~\ref{plan0}(b,d). Under certain restrictions it
can be shown using Rellich's lemma, originally proven in
\cite{Rellich43}, that the radiated field of a nonradiating source is
zero outside the source region \cite[Thm.~2.3]{Gbur03},
\cite[Lem.~1]{ColtonKress18}, \cite[p.~3824]{Blasten21},
\cite[Eq.~(3.4a)]{KimWolf86}, \cite[Thm.~1]{Vesalainen14}. In
contrast, Figure~\ref{plan0}(b) shows that $H^{\rm sc}(\myvec\rho)$ is
quite large between the walls but outside the source region. This
raises questions about the validity of applying Rellich's lemma to the
nonradiating sources that are induced by an incident wave.

\section{Conclusions}
\label{conclusion}

It is evident that an infinitely thin horizontal PEC wall is entirely
invisible to the incident wave~(10). This fundamental property gives
rise to some rather surprising results discussed in this paper. One
key outcome is that it is possible to design low-signature PEC
structures composed of two horizontal PEC walls, with either a PEC
fusiform cavity or a finite-periodic array of PEC cavities positioned
between them. Within these cavities, objects can be concealed from an
incident plane wave, effectively making the low-signature PEC
structures act as invisibility cloaks.

In addition to presenting the low-signature structures, the paper
highlights three main findings. First, the inverse problem of
determining the shape of a PEC surface does not necessarily have a
unique solution when only single-frequency incident waves are
used. Second, based on reciprocity, the far-field amplitude
$F^{\rm sc}(\theta)$ satisfies $F^{\rm sc}(0)\approx 0$ and
$F^{\rm sc}(\pi) \approx 0$ when the low-signature structures are
illuminated by any TM wave generated within $\Omega_{\rm rec}$. Third,
it is demonstrated that non-radiating sources can produce significant
electromagnetic fields near the source location, which contradicts an
established opinion regarding non-radiating sources.

\subsubsection*{Acknowledgments}

This work was supported by the Swedish Research Council under contract
2021-03720.

\vspace{15mm}
\appendix
\centerline{\LARGE{\bf Appendix}}
\section{Notes on the numerical scheme}
\label{app:A}

In~\cite{HelsJiang25}, a boundary integral equation (BIE) based
numerical scheme is presented for solving the exterior Neumann
problem~(\ref{PDE1},\ref{PDE2},\ref{PDE3}) and the corresponding
exterior Dirichlet problem. This scheme applies to structures with
boundaries $\Gamma$ being collections of open arcs, such as those
shown in Figures~\ref{combgeom} and~\ref{geometrireciproc}. It is used
for all computations in the present paper. In the context
of~(\ref{PDE1},\ref{PDE2},\ref{PDE3}) it contains steps such as:
choosing a layer-potential field representation for
$H^{\rm sc}(\myvec\rho)$; rewriting~(\ref{PDE1},\ref{PDE2},\ref{PDE3})
as a BIE based on that representation; Nystr{\"o}m discretization of
the BIE on a graded mesh, accelerated and stabilized with recursively
compressed inverse preconditioning (RCIP)~\cite{Tutorial}; and
iterative solution of large linear systems using the generalized
minimal residual method (GMRES), accelerated with the fast multipole
method (FMM)~\cite{fmm2d}.

We shall not review much detail of the scheme in~\cite{HelsJiang25}
here, but chiefly mention that, for the exterior Neumann problem, it
prescribes the choice of field representation
\begin{equation}
  H^{\rm sc}(\myvec\rho)=
  -2\int_{\Gamma}\frac{\partial\Phi_k}{\partial\nu'}(\myvec\rho,\myvec\rho')
  S_k\varrho(\myvec\rho')\,{\rm d}\ell'\,,
  \quad \myvec\rho\in\Omega\,.
\label{eq:rep2}
\end{equation}
Here $\varrho(\myvec\rho)$ is a layer density on $\Gamma$ to be
determined, $\Phi_k(\myvec\rho,\myvec\rho')$ is the fundamental
solution to the Helmholtz equation in the
plane~\cite[Eq.~(3.60)]{ColtonKress98}, $S_k$ is the single-layer
operator~\cite[Eq.~(3.8)]{ColtonKress98}, ${\rm d}\ell$ is an element
of arc length, and
$\partial/\partial\nu'=\nu(\myvec\rho')\cdot\nabla'$. Insertion
of~(\ref{eq:rep2}) into~(\ref{PDE2}) gives the BIE with composed
integral operators
\begin{equation}
  T_k(-S_k)\varrho(\myvec\rho)=
  -\nu\cdot \nabla H^{\rm in}(\myvec\rho)\,,
  \quad \myvec\rho\in\Gamma\,,
  \label{eq:main2}
\end{equation}
which is~\cite[Eq.~(31)]{HelsJiang25} and where $T_k$ is the
hypersingular operator~\cite[Eq.~(3.11)]{ColtonKress98}.

While the scheme in~\cite{HelsJiang25} with~(\ref{eq:rep2}) is
generally applicable to exterior Neumann problems, in the present
paper we only use~(\ref{eq:rep2}) for $\Gamma$ as in
Figure~\ref{geometrireciproc}. For $\Gamma$ as in
Figure~\ref{combgeom}, we take advantage of that $\Gamma_{\rm fc}$ is
a closed contour, rather than a general collection of open arcs, and
replace~(\ref{eq:rep2}) with
\begin{multline}
  H^{\rm sc}(\myvec\rho)=
  -2\int_{\Gamma_{\rm hw}}\frac{\partial\Phi_k}{\partial\nu'}
  (\myvec\rho,\myvec\rho')
  S_k\varrho(\myvec\rho')\,{\rm d}\ell'\\
  -2\int_{\Gamma_{\rm fc}}\Phi_k(\myvec\rho,\myvec\rho')\varrho(\myvec\rho')\,
  {\rm d}\ell'\,,
  \quad \myvec\rho\in\Omega\cup\Gamma_{\rm fc}\,,
\label{eq:rep1}
\end{multline}
which is simpler than~(\ref{eq:rep2}). Insertion of~(\ref{eq:rep1})
into~(\ref{PDE2}) gives a BIE, analogous to~(\ref{eq:main2}), which in
composed block operator form can be written
\begin{equation}
  \begin{bmatrix}
    T_k^{\rm (hw,hw)} & -K_k^{\rm A(hw,fc)} \\
    T_k^{\rm (fc,hw)} & I^{\rm (fc,fc)}-K_k^{\rm A(fc,fc)}
  \end{bmatrix}
  \begin{bmatrix}
    -S_k^{\rm (hw,hw)} & 0 \\
    0 & I^{\rm (fc,fc)}
  \end{bmatrix}
  \begin{bmatrix}
    \varrho_{\rm hw}(\myvec\rho)\\
    \varrho_{\rm fc}(\myvec\rho)
  \end{bmatrix}
  =
  \begin{bmatrix}
    g_{\rm hw}(\myvec\rho)\\
    g_{\rm fc}(\myvec\rho)
  \end{bmatrix}.
\label{eq:main1}
\end{equation}
Here $I^{(i,j)}$, $S_k^{(i,j)}$, $K_k^{{\rm A}(i,j)}$, $T_k^{(i,j)}$
are operators acting on densities at $\Gamma_j$ and evaluated at
$\Gamma_i$ with $i,j={\rm hw,fc}$. Furthermore, $I$ is the identity
operator, $K_k^{\rm A}$ is the adjoint double-layer
operator~\cite[Eq.~(3.10)]{ColtonKress98}, and
$\varrho_{\rm hw}(\myvec\rho)$, $\varrho_{\rm fc}(\myvec\rho)$,
$g_{\rm hw}(\myvec\rho)$, $g_{\rm fc}(\myvec\rho)$ are the
restrictions of $\varrho(\myvec\rho)$ and of
$-\nu\cdot\nabla H^{\rm in}(\myvec\rho)$ to $\Gamma_{\rm hw}$ and to
$\Gamma_{\rm fc}$, respectively.

We emphasize that the machinery for applying RCIP to~(\ref{eq:main2}),
described in~\cite{HelsJiang25}, also applies to~(\ref{eq:main1}) and
that the subsequent computational steps of the scheme are the same
regardless of whether~(\ref{eq:rep2}) and (\ref{eq:main2})
or~(\ref{eq:rep1}) and~(\ref{eq:main1}) are chosen.

\end{document}